# Radiation Skyshine Calculation with MARS15 for the mu2e Experiment at Fermilab

**A.F. Leveling**
Fermi National Accelerator Laboratory

## Abstract

The Fermilab Antiproton source is to be repurposed to provide an 8 kW proton beam to the Mu2e experiment by 1/3 integer, slow resonant extraction. Shielding provided by the existing facility must be supplemented with in-tunnel shielding to limit the radiation effective dose rate above the shield in the AP30 service building. In addition to the nominal radiation shield calculations, radiation skyshine calculations were required to ensure compliance with Fermilab Radiological Control Manual. A complete model of the slow resonant extraction system including magnets, electrostatic septa, magnetic fields, tunnel enclosure with shield, and a nearby exit stairway are included in the model. The skyshine model extends above the beam enclosure surface to 10 km vertically and 5 km radially.

**Facility Overview**

The dominant source of radiation dose during Mu2e operation is the Delivery Ring extraction system, located in the beam enclosure below the AP30 service building. The AP30 Anti-proton source service building was originally designed in conjunction with the Accumulator/Debuncher Rings for a mW power, secondary anti-proton beam and could be operated nominally at up to 13 watts of 8 GeV primary proton beam. The shield between the beam tunnel and service building is 10 feet thick (3.048 m). The Anti-proton source Debuncher Ring (now, the Delivery Ring) is being reconfigured to condition and extract an 8 kW, 8 GeV proton beam by 1/3 integer, slow resonant extraction, a relatively lossy process. If this facility was to be built in a "green field", a shielding thickness of 18 to 22 feet might be chosen. To compensate for the shielding deficit, an in-tunnel steel shielding system has been designed. A MARS model of the existing facility was created which includes a portion of the Delivery Ring, the slow resonant extraction system, extration beam line, and a nearby exit stairway. A longitudinal elevation view of the facility model is shown in Figure 1 while transverse elevations views which illustrate details of the exit stairway are shown in Figure 2.

**Figure 1: A longitudinal elevation view of the MARS model used in this work is shown. The shielding berm increases to 13 feet (3.96 m) just downstream of the indicated stairway. The AP 30 Service Building walls and roof are not shown in the figure.**

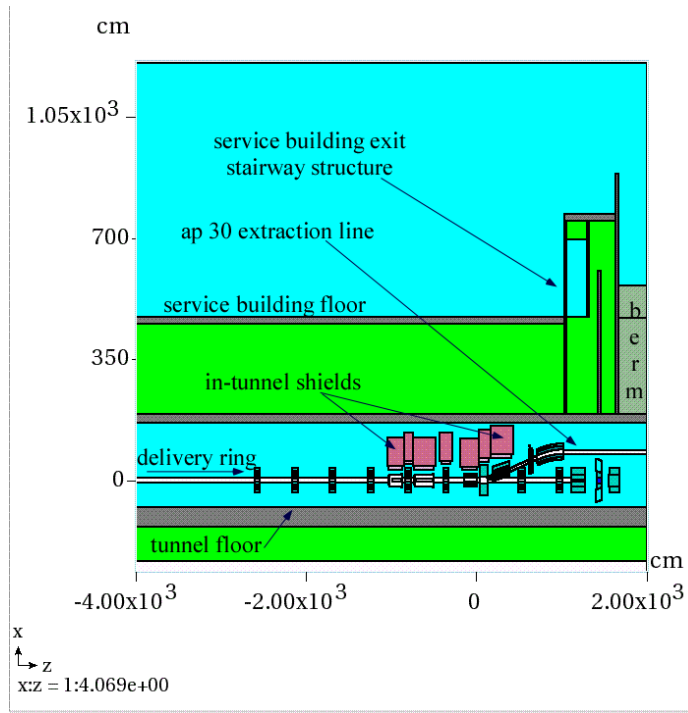

**Figure 2: The shielding details of the exit stairway**

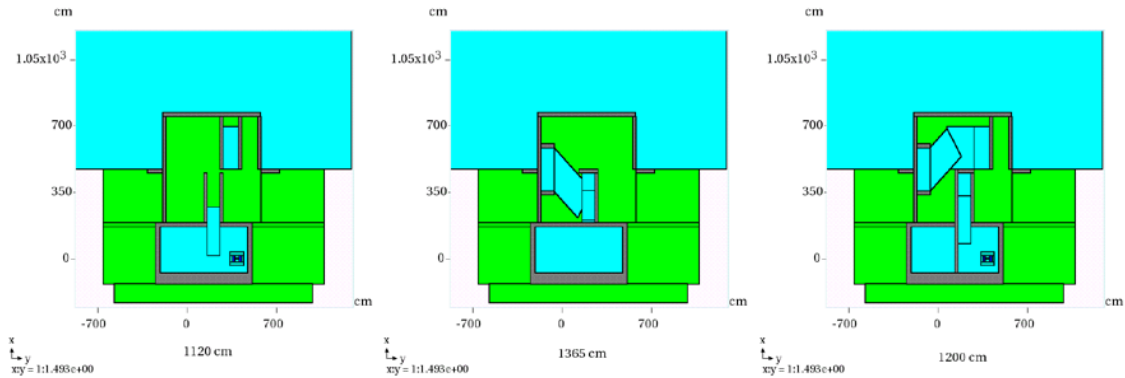

## Extraction System Configuration and Alignment

Details of the MARS model for the extraction system including a portion of the Delivery Ring, the electrostatic septa, various quadrupoles, extraction Lambertson magnet, C magnet, and vertical bending magnet are shown in Figure 3. The model includes magnetic fields in the quadrupole and bending magnets. The electric fields of the electrostatic septa are approximated by a magnetic field vector at 90 degrees to the nominal electric field. A total of 850 graphite and tungsten foils are included in the model.

**Figure 3: The Delivery Ring and extraction system details are shown here in a series of expanded views.  Horizontal scales for the images are left, 20 m; middle, 6 m; and right, 0.06 m**

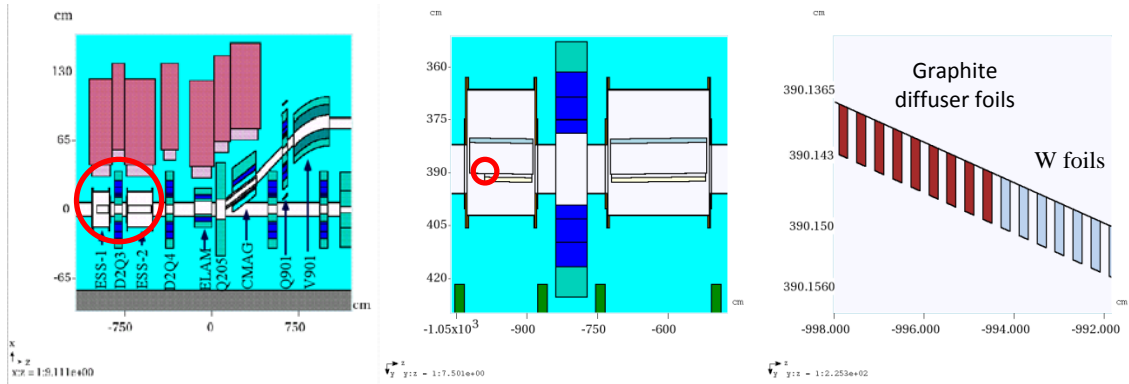

The 2 m long Lambertson magnet (ELAM in Figure 3) is the limiting aperture in the extraction system. Cross section views of the circulating and extraction apertures are illustrated in Figure 4. The circulating beam in the Delivery Ring passes through the non-field region of the Lambertson magnet while the extracted beam passes through the field region where it is deflected vertically upward. A part of the beam which intersects the foil plane scatters and is lost in the Lambertson magnet at the septum, a 3 mm wide steel divide between the circulating and extraction channels. The septum is acts as a magnetic field flux return; consequently, its design thickness is a constraint which results in unavoidable beam loss.

**Figure 4: The beam position at the upstream and downstream ends of the Lambertson magnet apertures can be seen in the figure at left. Beam scattered by the electrostatic septa foils is stopped in the steel in the region depicted by the circle in the right image.**

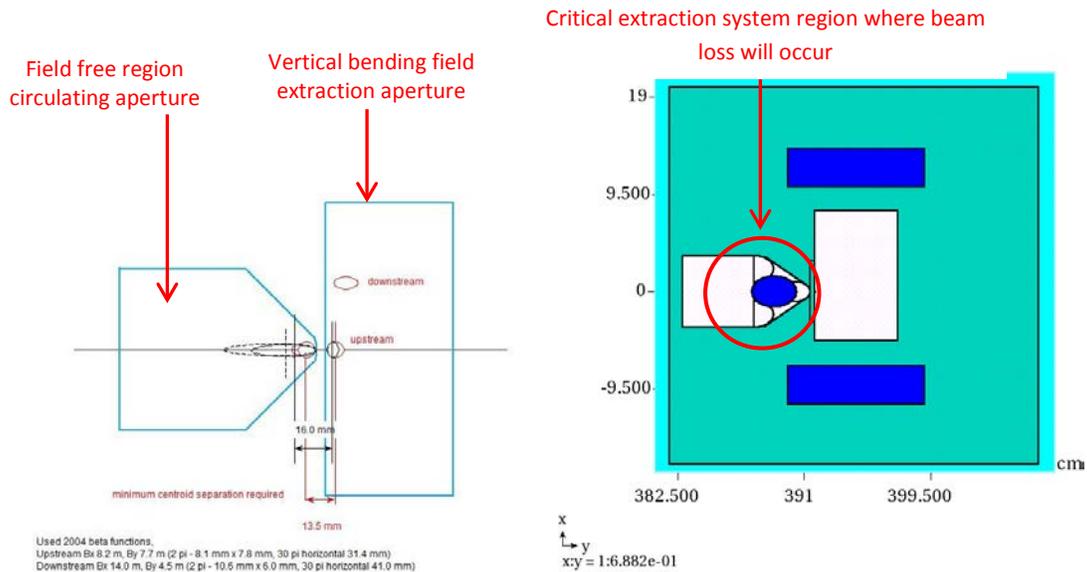

In the MARS model, alignment of extraction system components is necessary to optimize beam transmission and minimize beam loss. Scattering of the proton beam incident on the electrostatic septa foil planes is unavoidable. Losses from scattering in the foils are indistinguishable from losses due to misalignment of extraction system magnets. Therefore, to aid in the alignment process, the wire plane foils are temporarily treated as black holes. This permits the positioning of the extraction Lambertson, C-magnet, and other extraction line components to minimize beam loss. Once loss-free extraction positions are determined, the foil planes are returned to their normal material properties to establish conditions for normal beam loss. Surface detectors were also included in the model to determine the fraction of the beam lost in the tracking studies shown in Figure 5. The total beam loss for the region is estimated to be 1.25% or about 100 watts.

**Figure 5: Extraction system component alignment is illustrated in these figures. At left, the electrostatic septa foil planes are treated as black holes. The separated beams scrape on the misplaced Lambertson magnet resulting in a particle shower indicated by the green neutron traces. The Lambertson magnet is then repositioned so that the separated beams pass cleanly through the circulating and extraction orbits as shown at right.**

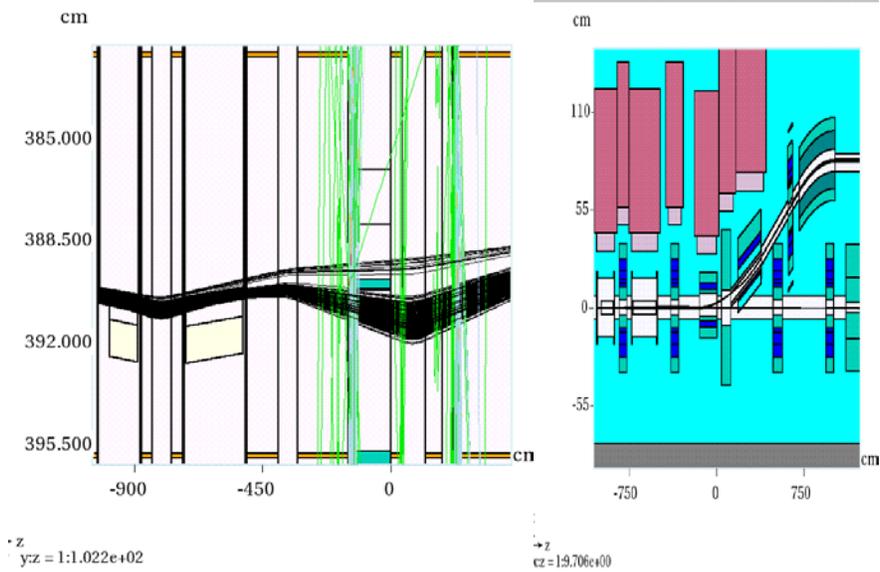

## Supplemental Shielding System

The AP30 service building shielding is limited to 10 feet (3.048 m). Various schemes have been examined to supplement the shielding externally but all options explored were found to be either impracticable or cost prohibitive. An in-tunnel, supplemental shielding system was devised as an alternative. The modular shield design can be adopted on a location by location basis as required. The design features of such a system are illustrated in Figure 6. Seven of the supplemental systems were incorporated in the MARS model to shield extraction beam losses resulting from the electrostatic septa, Lambertson magnet, C-magnet, and three quadrupoles in Figure 3 and Figure 5.

**Figure 6: Supplemental in-tunnel shielding is shown above the Lambertson magnet in the figure.**

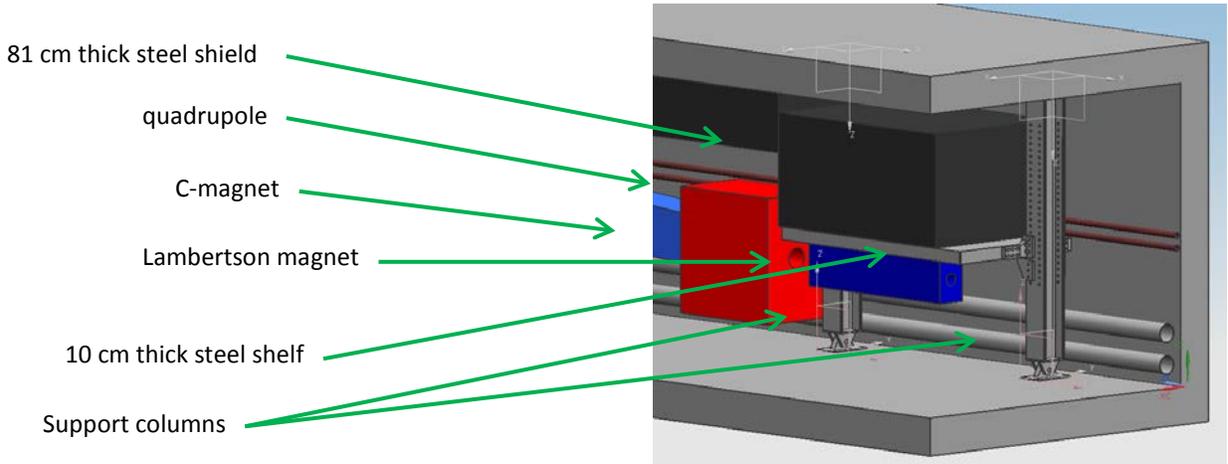

In addition to the steel shield shown in Figure 6, a composite concrete/marble shield is to be located in the aisle of the tunnel adjacent to the extraction devices. Residual radiation dose rates due to the unavoidable 100 watt extraction beam loss will be shielded to limit worker exposure during Delivery Ring maintenance periods. An illustration of the aisle shield is shown in Figure 7.

**Figure 7: A concrete/marble shield is to be placed in the tunnel adjacent to extraction system components to limit worker radiation exposure during maintenance periods. The shields are mounted on movable platforms to facilitate access to extraction system components.**

Aisle way shield is required to limit worker exposure during
maintenance activities
45 cm concrete
10 cm marble (not shown)

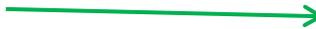
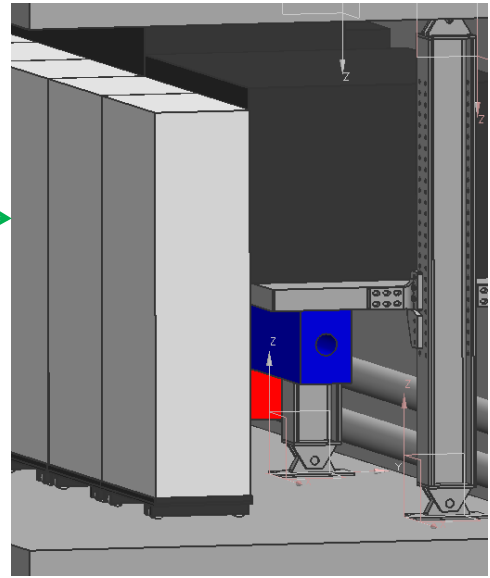

## Radiation Skyshine Model

The radiation skyshine model eventually developed for this work is a cylindrical volume with a radius of 5 km and a height of 10 km. Initially, the goal in making the model was to understand radiation effective dose at a radial distance of 500 meters and a height of up to 500 meters. As work on the calculation proceeded, and because grid computing resources were being employed, it became apparent that a significantly larger atmospheric model could be used without the need for extraordinary calendar time to complete the calculation. Therefore, the dimensions of the model were increased.

The base of the model is a concrete disk 2 m thick and 5 km radius. A tissue equivalent (TE) layer of detector, 0.3 m in height, covers the concrete disk. For the first 50 m, the TE layer is subdivided into 1 meter radial bins. From 50 m to 5,000 m, the TE disk is divided into 10 m radial bins. A MARS histogram volume (air), 100 m long by 100 m wide by 1.7 m high was placed in the atmosphere just above the TE detector at the model center. The purpose of this histogram is to determine radiation effective dose rate due to direct and skyshine sources in the service building, the adjacent parking lot, and the nearby service road. Details of the model are shown in Figure 8 and Figure 9.

The placement of the particle source between the TE detector and histogram volumes was intentional. The source propagates upward through the histogram volume which gives a measure of the direct component. Reflected sky shine passes downward through the histogram volume and provides a pure skyshine component in the TE detector.

**Figure 8: Skyshine model feature for radius = 500 m and height = 10 m**

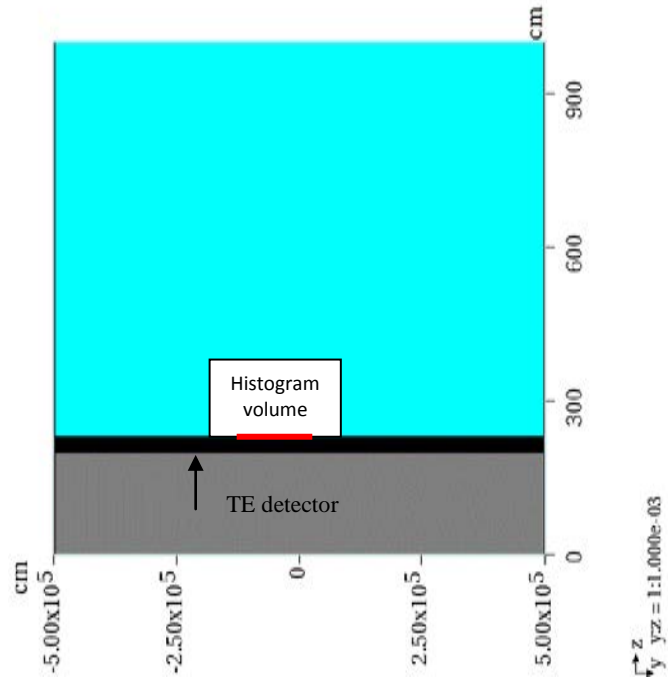

**Figure 9: A cross section of the full model is shown in the left figure while the plan view of the model through the TE detector for the first 50 m in radius in shown at right.**

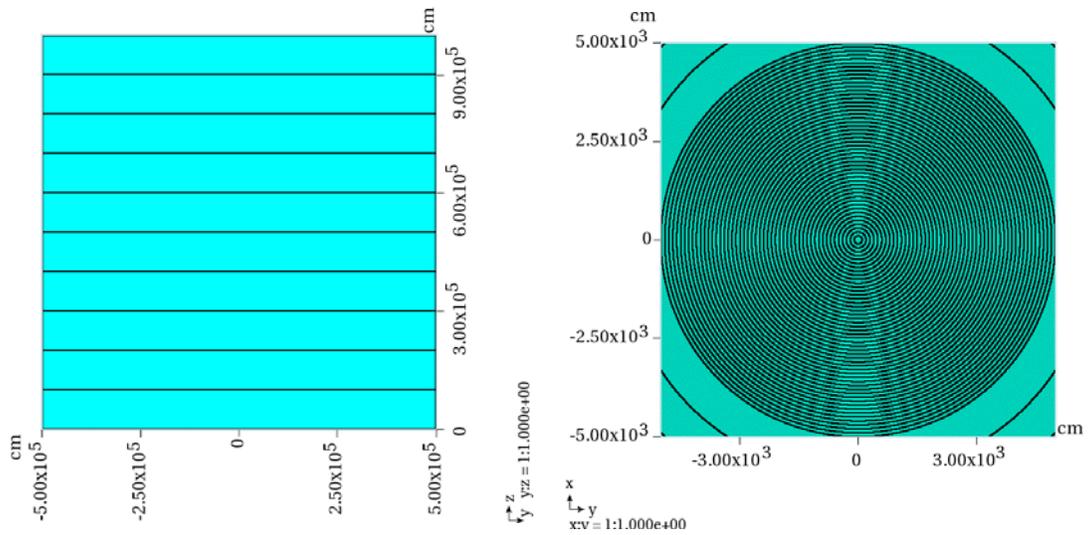

The density of the atmosphere as a function of height was calculated using the NASA earth atmosphere model for the troposphere for altitudes less than 11,000 m [4]. The temperature corrected density was calculated at the elevation for the center of each of ten 1 km layers (local ground elevation at the AP30 service building is 744 feet) and assumed to be constant throughout each layer. The atmosphere was modelled with weight fractions of the following elements: nitrogen (0.746), oxygen (0.24), argon (0.013), and hydrogen (0.001). The density of the atmosphere was found to have a profound effect on the shape of the plume.

Wilson Hall, the 16 story central laboratory building, is sited approximately 500 m from the AP30 service building. This building would be exposed to radiation directly emitted from the surface of the AP30 service building floor. A second MARS skyshine model, referred to as the DIRECT model, was also employed to determine radiation effective dose rate as a function of floor at Wilson Hall, due to direct and skyshine radiation sources. The model, created in root geometry, consists of at TE cylindrical shell centered on the AP30 service building with a radius of 500 m and a height of 70 m corresponding to the height of Wilson Hall. The TE cylinder was subdivided into 16 layers representing the approximate division of floors within Wilson Hall. Each of the layers was subdivided into 10 degree bins in azimuth. The details of the model are illustrated in Figure 10.

**Figure 10: The blue arrow points to the approximate location of the slow resonant extraction system in the AP30 service building. The tail of the arrow is directed toward Wilson Hall, an angle of 23 degrees relative to the direction of the incident beam. The figure at right shows a representation of the DIRECT model. The layers represent approximate floor locations at Wilson Hall and are further subdivided into 36 angular bins of 10 degrees azimuth. The cylindrical shell is centered at the AP30 service building.**

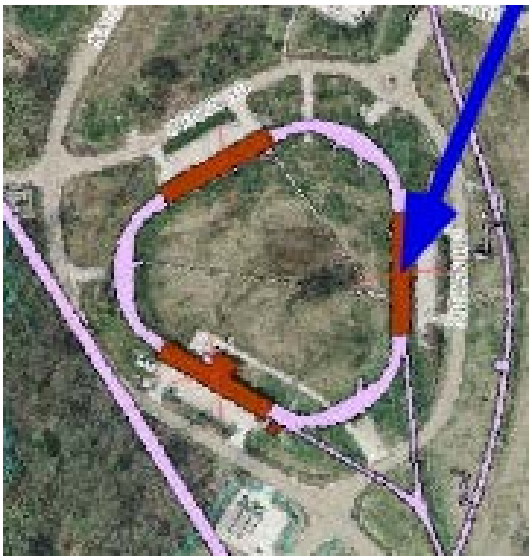 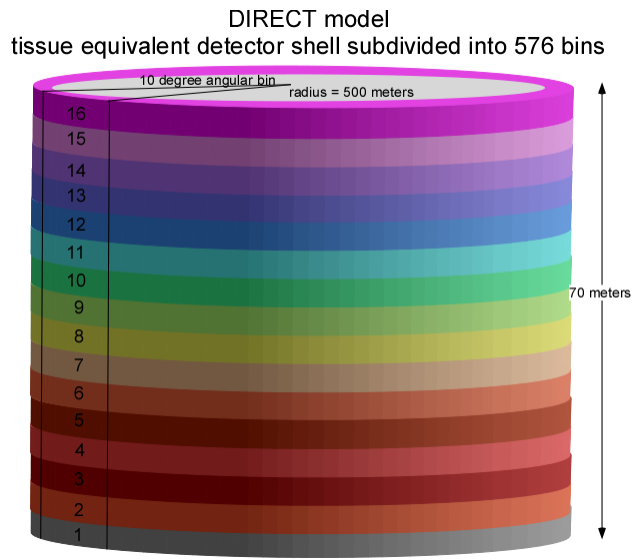

## MARS Simulations

A series of four MARS simulations was required for this work. In the first run (called stage 1), the 8 GeV proton beam is introduced to the slow resonant extraction system. The resulting shower is propagated through the slow resonant extraction system, the in-tunnel shielding system, and to a surface just outside of the tunnel. The goal of the first run is to write a file of shower particles at a surface defined outside of the tunnel containing the slow resonant extraction system. The particle file consisting of $2\times10^5$ to $1\times10^6$ particles contains the particle type, the weight, energy, positions in x, y, and z and the direction cosines. In the second run (called stage 2), the stage 1 particle file is used as a source term to continue propagation of the shower through the remaining shield above the tunnel. A second set of surfaces, the service building floor and the top surface of the stairway structure, were established to collect a new set of shower particles for the stage 2 run. The stage 2 run particle files were used as source terms for the third (skyshine) and fourth (direct) calculations. Figure 11 shows the location in elevation at which the particle source files were written. Histograms indicating the total flux during the stage 1 and stage 2 runs are shown in Figure 12.

**Figure 11: Elevation view of AP 30 service building showing elevations at which stage 1 and stage 2 particle showers were collected.**

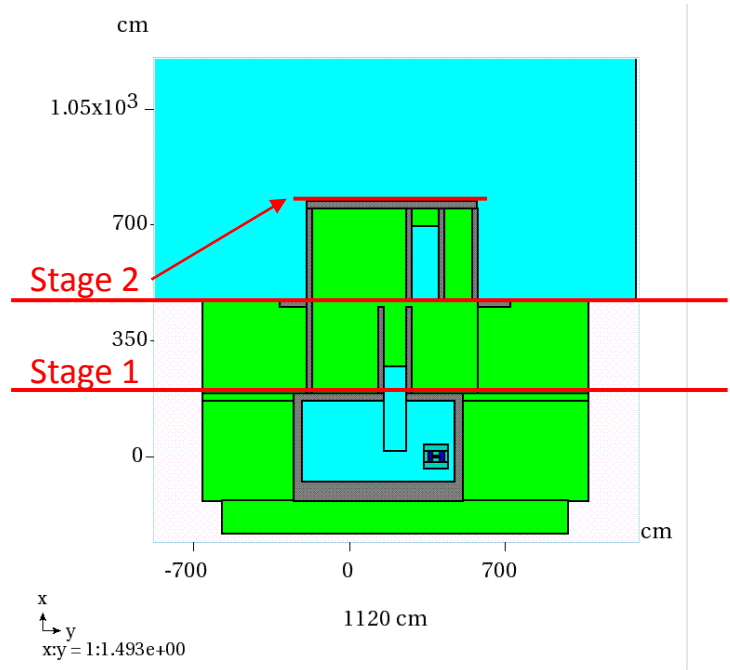

**Figure 12: Longitudinal elevation views in the plane of the proton beam are shown in these images. Histograms of total flux created during the stage 1 (left) and stage 2 (right) runs are indicated. The yellow lines indicate particle collection surfaces for the two runs.**

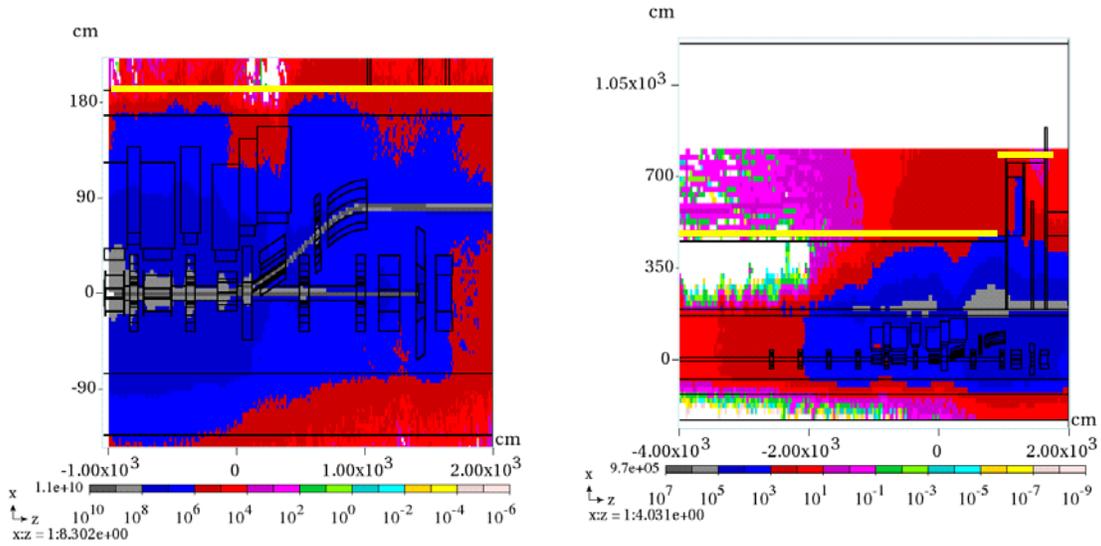

## MARS Skyshine and Direct Results

Histograms in elevation view of total effective dose rate for the 8 kW proton beam extraction with a 1.25% beam loss are shown in Figure 13. The plume in the xz plane is tipped due to the forward momentum tendency of the emerging particle shower. The plume in the yz plane is tipped to beam left due to the proximity of the beam transport system adjacent to the tunnel wall at beam right; i.e., the tunnel wall suppresses the plume at beam right.

**Figure 13: The prompt total effective dose rate in mrem/hr is shown for the xz plane at left and the yz plane at right.**

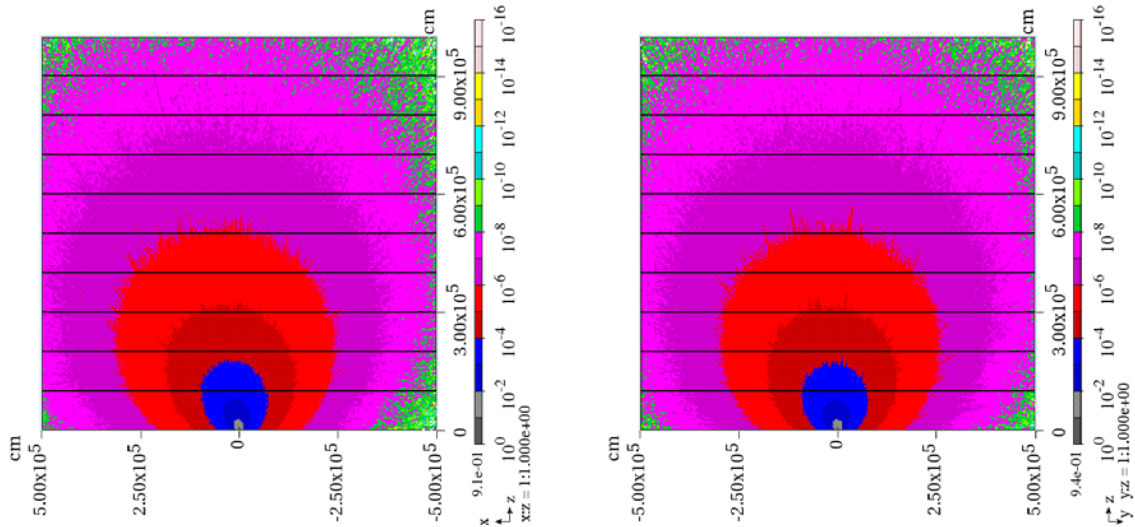

Histograms in a plan view of total effective dose rate for the 8 kW proton beam extraction with a 1.25% beam loss are shown in Figure 14. The non-symmetric nature of the plume is further amplified in these images.

**Figure 14: Plan view of average total effective dose rate in mrem/hr for ten 1 km layers of atmosphere above the AP30 service building**

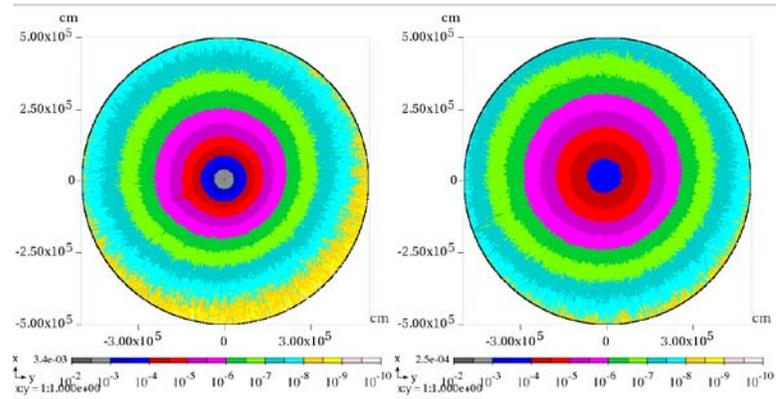

Layer 1 & 2

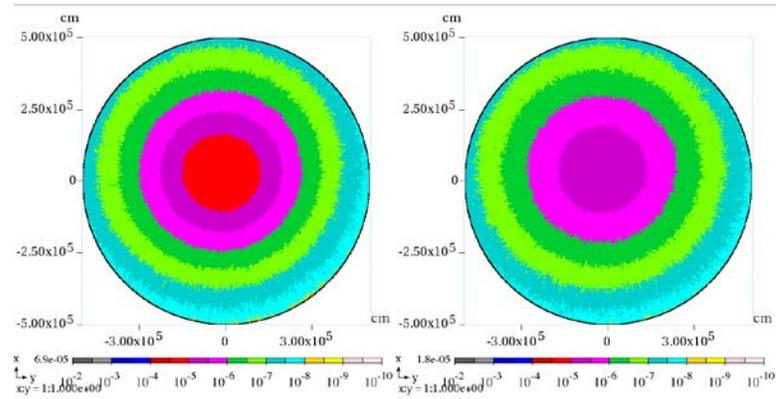

Layer 3 & 4

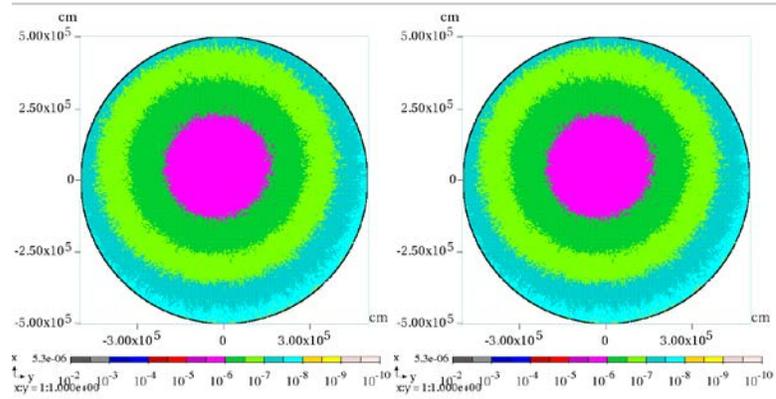

Layer 5 & 6

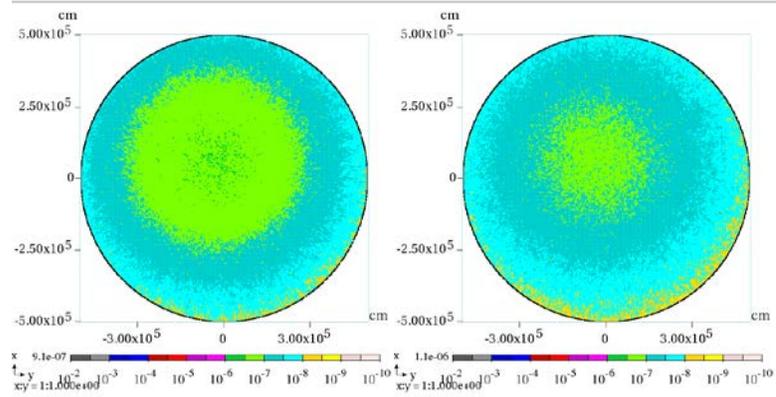

Layer 7 & 8

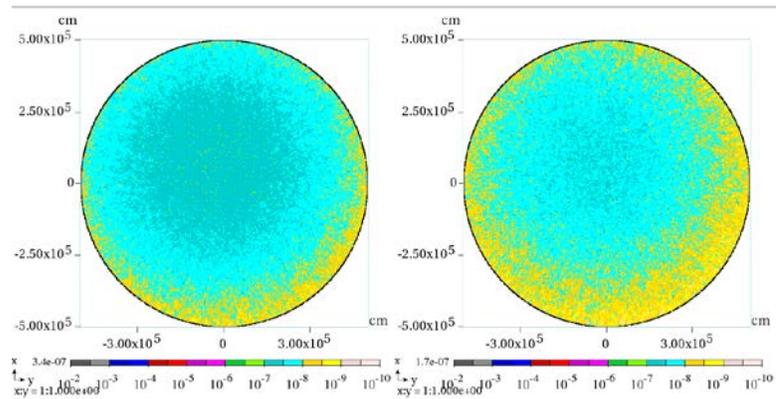

Layer 9 & 10

The result of total effective dose rate in the 5 km radius TE detector is shown in Figure 15. The dose rate at Wilson Hall (500 m) is 0.17 mrem per year

The result for prompt effective dose rate in the volume histogram in the vicinity of the AP30 service building is shown in Figure 16.

**Figure 15: The annual radiation effective dose rate for continuous Mu2e operation in the 5 km TE detector is shown in blue while statistical errors are shown in red as a function of distance from the center of the model.**

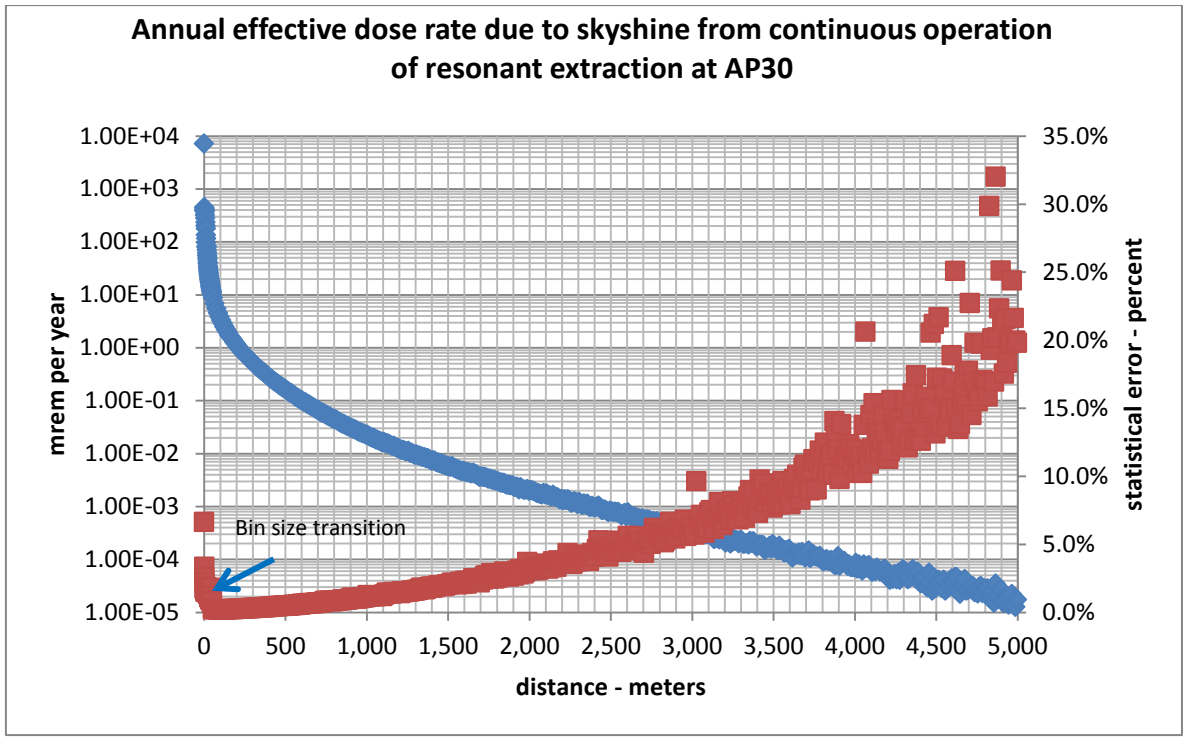

**Figure 16: 1: Histogram of prompt effective dose rate in mrem/hr for slow resonant extraction of 8 GeV, 8 kW beam loss with 1.25% beam loss. Enumerated lines legend is: 1. Delivery Ring centerline; 2: Tunnel outer concrete edge; 3. Edge of service building; 3/4: Parking Lot; and 5: edge of Indian Road**

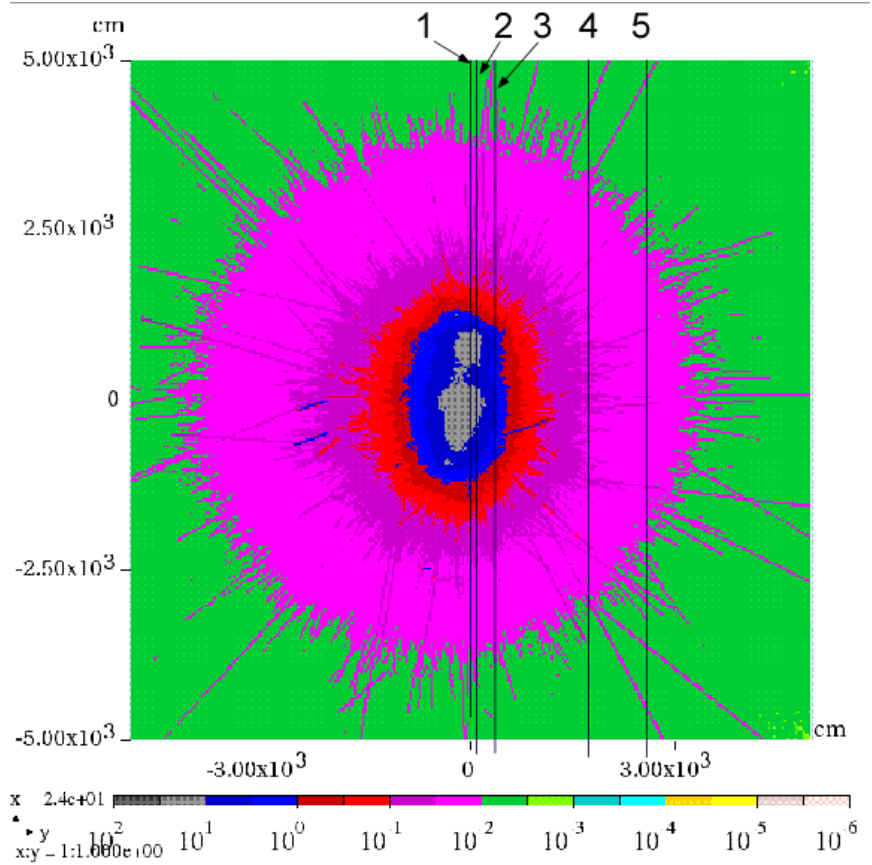

Finally, the result of the direct calculation, which includes direct and skyshine sources as a function of floor in Wilson Hall, corresponding to an angle of 23 degrees in azimuth relative to the forward incident beam direction, is shown in Figure 17.The 1[st] floor does not receive a direct contribution from the AP30 service building since the areas share a common elevation. The direct effective dose rate contribution at a given floor can be approximated by subtracting the 1st floor rate from the combined rate. Wilson Hall is a massive concrete structure. No credit is taken for the shielding provided by the building. Consequently, the calculations are conservative except perhaps where offices are located in the glass-walled crossovers at the south face of Wilson Hall.

**Figure 17: The combined skyshine/direct effective dose rate as a function of floor elevation in Wilson Hall is shown at the arrow in the plot. The difference between the 1st floor and other floor effective dose rates is due to the direct effective dose rate.**

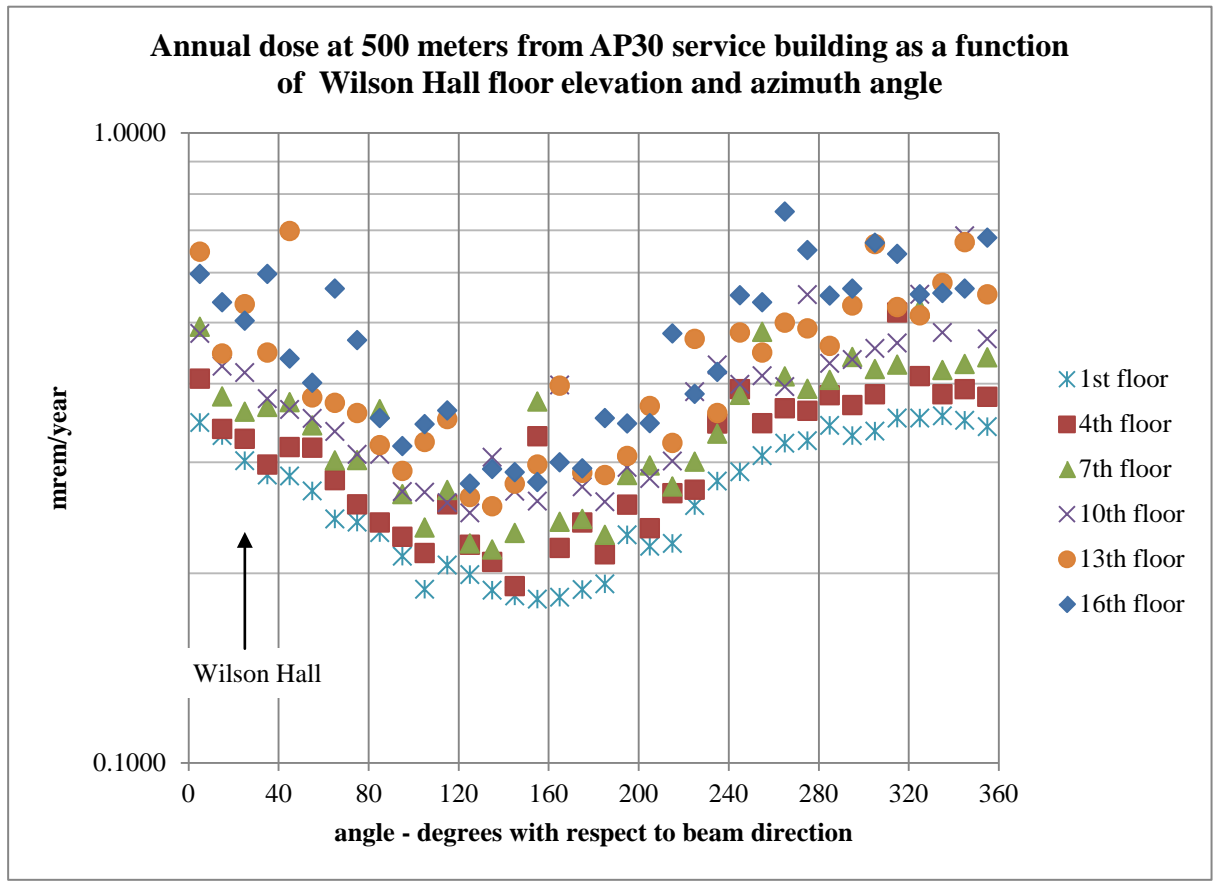

**Conclusions**

A model of the resonant extraction system has been created in which 1.25% beam losses is realistically distributed. A supplemental shield system design has been produced. The resulting calculated skyshine and direct effective dose rates fall within all limitations of the Fermilab Radiological Controls Manual. However, an active protection system will be required to limit radiation effective dose rates significantly higher than those calculated in the work reported here. Additional sources of beam loss at the antiproton source facilities could lead to additional sources of skyshine. Those additional sources must be included with the results reported here when/if they are observed. Based upon the beam loss scenario considered here, it will be prudent to exclude personnel access to the AP30 service building during mu2e beam operations.

**Acknowledgements**

Consultation with a rather large group of people was required to produce this work. First and foremost, the author acknowledges the contributions by Don Cossairt and Nikolai Mokhov for insights into the skyshine problem and for application of the MARS code to solve it.

Vladimir Nagaslaev provided the incident beam parameters for the resonant extraction system as well as the layout of the ESS septa placement and angles. Jim Morgan provided the design layout of the extraction system along with beam line coordinates and magnet currents. Carol Johnstone provided starting alignment data for the extraction line components. Corey Crowley designed the modular in-tunnel shielding system. Brian Hartsell provided the preliminary electrostatic septa mechanical design adapted for the model. Mike Vincent provided the C-magnet geometry. Rob Kutschke and Andrei Gaponenko made advances in MARS grid job submission procedures and taught the author to use them. Steve Werkema provided general help and consul in the development and execution of this model and the many previous versions. He also reviewed this paper and is partly responsible for whatever clarity of thought it may provide. Finally, the calculations used to produce results presented here required many years of cpu time. This work would not have been possible without the support of the Fermilab grid computing system; 6 years ago, such work would not have been possible.